\documentstyle[psfig,referee]{mn}

\def\gte{\,\lower.6ex\hbox{$\buildrel >\over \sim$} \, }
\def\lte{\,\lower.6ex\hbox{$\buildrel <\over \sim$} \, }

\title{Why do starless cores appear more flattened than protostellar cores?}

\author[S.\,P.\,Goodwin, D.\,Ward-Thompson \& 
A.\,P.\,Whitworth]{S.\,P.\,Goodwin, D.\,Ward-Thompson 
\& A.\,P.\,Whitworth \\ Dept. of Physics \& Astronomy, 
Cardiff University, 5 The Parade, Cardiff, CF2 3YB, Wales, UK}

\date{}

\begin{document}

\maketitle

\begin{abstract}

We evaluate the intrinsic three dimensional shapes of molecular cores,
by analysing their projected shapes. We use the recent catalogue of
molecular line observations of Jijina et al. and model the data by the
method originally devised for elliptical galaxies.  Our analysis
broadly supports the  conclusion of Jones et al. that molecular cores
are better represented  by triaxial intrinsic shapes (ellipsoids) than
biaxial  intrinsic shapes (spheroids). However, we find that the best
fit to all of the data is obtained with more extreme axial ratios
($1:0.8:0.4$) than those derived by Jones et al.

More surprisingly, we find that starless cores have more extreme
axial ratios than protostellar cores -- starless cores appear more
`flattened'. This is the opposite of what  would be expected from
modeling the freefall collapse of  triaxial ellipsoids. The collapse
of starless cores would be expected to proceed most swiftly along the
shortest axis -- as has been predicted by every modeller since
Zel'dovich -- which should produce more flattened cores around
protostars, the opposite of what is seen.

\end{abstract}


\section{Introduction}

Stars form in dense interstellar regions known as molecular clouds,
and particularly in the densest cores of molecular clouds.  A great
deal of observational work  on molecular cloud cores has been carried
out and a large number of sites of star formation have been observed
in various molecular transitions (e.g. Myers, Linke \& Benson 1983;
Myers \& Benson 1983;  Benson \& Myers 1989).  The chief transitions
used tend to be various isotopes of CO (a tracer of column density)
and NH$_3$ (a tracer of volume density). A recent paper (Jijina, Myers
\& Adams 1999) cataloged all of the known molecular observations of a
large sample of nearby star-forming regions, listing many  of their
physical properties such as density, temperature, size and shape.
Such a catalogue provides a wealth of detailed information about the
star formation process. In this letter we concentrate on what we can
learn from this catalogue about the shapes of star-forming cores.

When observing molecular cloud cores, all we see are their projected
shapes  on the sky, whilst for detailed comparison with theory we
would like  information on their three dimensional shapes. Jijina et
al.  (1999) used their molecular-line maps of  molecular cores and
estimated the projected axial ratios ($q$) by  fitting elliptical
envelopes to the data.  Here we analyze the resulting distribution  of
projected axial ratios, on the assumption that molecular cores are
ellipsoidal, with a well-defined distribution of intrinsic axial
ratios ($1:\zeta:\eta$), and that the cores in the Jijina et
al. sample  are randomly oriented. Our aim is to constrain the
distribution of  intrinsic three-dimensional axial ratios.

\section{Deprojecting ellipticity}

A triaxial object, when viewed from any angle, appears in projection
as an ellipse.  Binney (1985) calculated the projected axial ratio $q$
($=b/a \lte 1$) for an ellipsoid having axial ratios $1:\zeta:\eta$
(where $\eta \leq \zeta \leq 1$) and viewed from an angle
$(\theta,\phi)$  as

\begin{equation}
q = \sqrt{ \frac{A + C - \sqrt{[(A-C)^2 + B^2]}}{A + C +
\sqrt{[(A-C)^2  + B^2]}}} ,
\end{equation}

\noindent where $A$, $B$ and $C$ are functions of the intrinsic axial
ratios  and the viewing angles given by

\[
A = \frac{{\rm cos}^2(\theta)}{\eta^2} \left[ {\rm sin}^2(\phi) +
\frac{{\rm cos}^2(\phi)}{\zeta^2} \right] +  \frac{{\rm
sin}^2(\theta)}{\zeta^2} ,
\]

\[
B = {\rm cos} (\theta) {\rm sin} (2\phi)  \left[ 1 - \frac{1}{\zeta^2}
\right] \frac{1}{\eta^2}
\]

\noindent and

\[
C = \left[ \frac{{\rm sin}^2(\phi)}{\zeta^2}  + {\rm cos}^2(\phi)
\right] \frac{1}{\eta^2} .
\]

The simplest statistical approach to take in determining whether a
given  distribution of projected axial ratios ($q$)  is drawn from a
single  distribution of intrinsic axial ratios ($\zeta,\eta$), is to
use an  assumed distribution of $\zeta$- and $\eta$-values to generate
a large  number (more than $10^5$) of projected $q$-values in a Monte
Carlo  simulation. These Monte Carlo $q$-values can be converted into
a  distribution function, and this `artificial' distribution function
can then be compared with the `real' (i.e. observed) distribution
function from the Jijina et al. data set. The KS-test and $\chi^2$
test can be used to evaluate the likelihood  of the assumed fit. 

Following Jones, Basu \& Dubinski (2001), we assume that the axial
ratios $\zeta$ and $\eta$ have Gaussian distributions characterized
by means ($\bar{\zeta},\bar{\eta}$) and standard deviations
($\sigma_\zeta,\sigma_\eta$), with the additional constraint that
$\zeta$  and $\eta$ must fall in the interval from 0 to 1.

\section{Results}

We find that the best fit to the observed $q$-values of the entire
Jijina et al. (1999) sample is obtained with $\bar{\zeta} \simeq
0.8\,$,$\sigma_\zeta \simeq 0.1$, $\bar{\eta} \simeq 0.4$ and
$\sigma_\eta \simeq 0.2\,$. It has $\chi^2$ probability of 0.74. In
contrast, Jones et al. (2001) find $\bar{\zeta} \simeq
0.9\,$,$\bar{\eta} \simeq 0.5$ and $\sigma_\zeta \simeq \sigma_\eta
\simeq 0.1\,$.  This difference is significant at the high ellipticity
end of the distribution (but not in the heart of the distribution as
both are good fits to the data).  To test this $10^5$ ellipticities
were drawn from both our best fit and that of Jones et al. and the KS 
test rejects the two distributions as being the same at very high 
probability ($>95$ per cent).  The rejection is at even higher
significance for less than $10^5$ samples.

The reason why our analysis rejects the Jones et al. (2001) best fit 
is that the  smallest  observed  axial ratio  $q = 0.175$
requires the combination of an extremely small short axis
($\sim3\sigma_\eta$ below $\bar{\eta}$) and a viewing direction within
a few degrees  of the mid-axis.  This is  extremely unlikely. Moreover
it should be  noted that the $q = 0.175$ core is not  a lone outlier;
there are four cores with $q < 0.25$.

Jijina et al. (1999) find significant differences in many of the 
cores' properties between starless cores (ones with no associated 
IRAS source) and protostellar cores (ones with at least one 
associated IRAS source). They also find significant differences 
between cores which are associated with star clusters and those which 
are not. We therefore repeated our analysis for these subsamples. 
Association with a star cluster does not appear to affect the 
distribution of intrinsic axial ratios. 

However, there does appear to be a statistically significant
difference  between starless cores and protostellar cores.  There are
79 starless cores and 179 protostellar cores in the sample, excluding
cores defined by Jijina et al. (1999) as being multiples, due to the
difficulties of correctly determining the projected axial ratios of
such cores.

Figure 1 shows the best fit for starless cores, obtained with 
$\bar{\zeta} = 0.8\,$, $\sigma_\zeta = 0.1\,$, $\bar{\eta} = 0.3\,$, 
and $\sigma_\eta = 0.2\,$.  This fit maximises the KS probability at
0.46 (with a $\chi^2$ probability of $0.85\,$), a reasonable fit to
the data. 

\begin{figure}
\centerline{\psfig{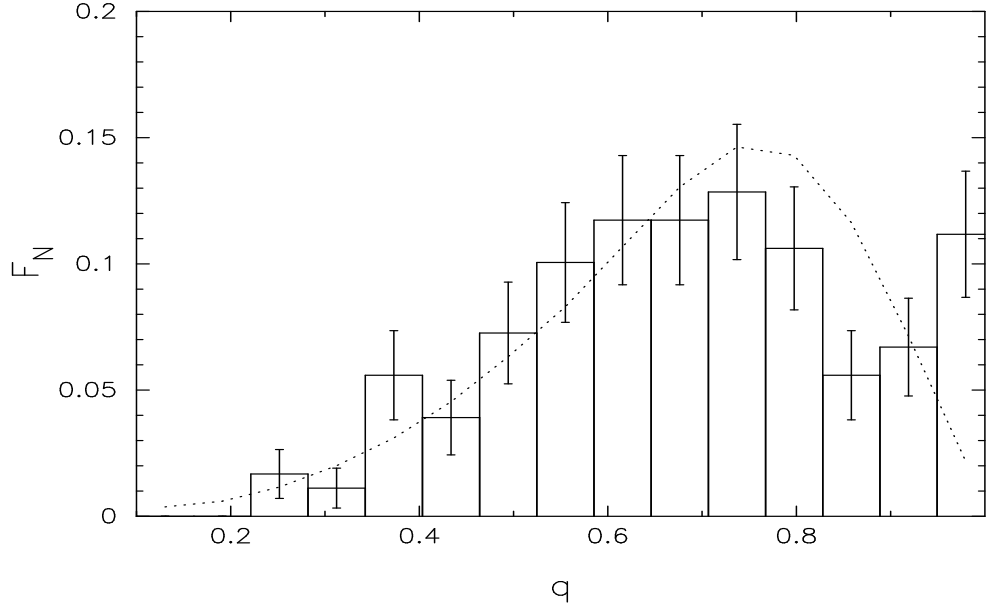}}
\caption{The best fit to the sample of 79 starless cores with axial 
ratios $1:0.8:0.3$ with 1 $\sigma$ spreads of 0.1 on the mid axis and 
0.2 on the short axis.  The bars show the $\sqrt{n}$ errors for each 
bin.}
\label{fig:starless}
\end{figure}

Figure 2 shows the best fit for the protostellar cores, obtained 
with $\bar{\zeta} = 0.8\,$, $\sigma_\zeta = 0.1\,$, $\bar{\eta} =
0.5\,$, and $\sigma_\eta = 0.2\,$. This fit has a KS probability of
only 0.01 (but a $\chi^2$ probability of $0.83\,$).  Whilst this is the
best fit it appears that a distribution formed so simply is not good
at representing the observed distribution.  The results are 
summarised in table 1.

\begin{figure}
\centerline{\psfig{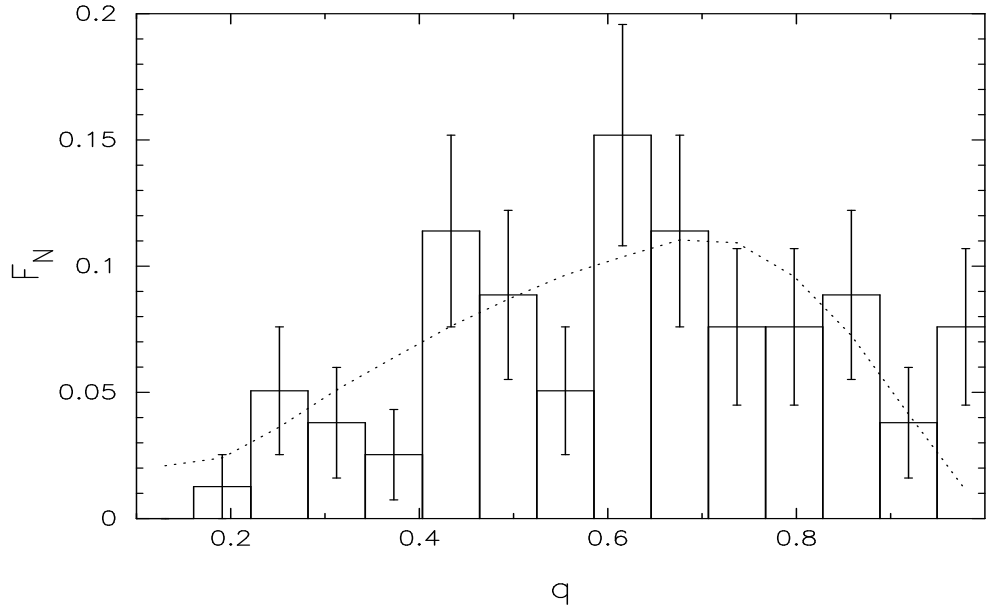}}
\caption{The best fit to the sample of 179 protostellar cores  with
axial ratios $1:0.8:0.5$ with 1 $\sigma$ spreads of 0.1  on the mid
axis and 0.2 on the short axis.  Again the bars show the $\sqrt{n}$
errors for each bin.}
\label{fig:protofit}
\end{figure}

Figure 3 demonstrates the fact that starless and protostellar cores
have significantly different distributions, by comparing their
cumulative distribution functions with our average fit obtained using
$\bar{\zeta} = 0.8\,$, $\sigma_\zeta = 0.1\,$, $\bar{\eta} = 0.4\,$,
and  $\sigma_\eta = 0.2\,$. Evidently neither data set is matched
especially well by  this average fit. Instead, the average fit lies 
comfortably between the two  cumulative distributions, with the majority of
starless cores lying above the average fit, and the majority of
protostellar cores lying below it.  This difference is significant as
the KS test shows that starless and protostellar cores only have a 5
per cent probability of being drawn from the same distribution.

\begin{figure}
\centerline{\psfig{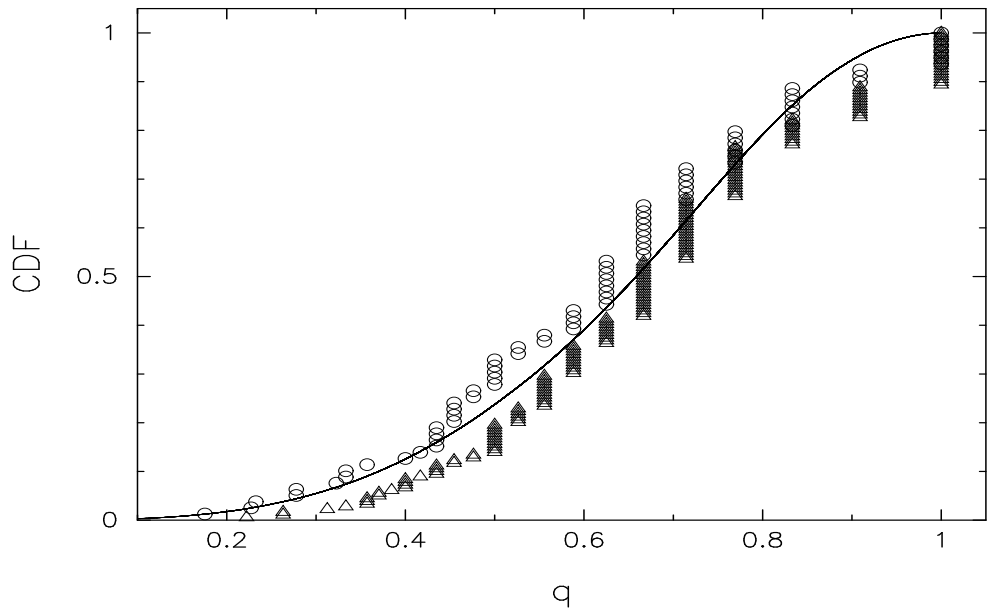}}
\caption{The cumulative distribution functions of starless cores (open
circles) and protostellar cores (filled triangles) and an `average'
fit of $1:0.8 \pm 0.1: 0.4 \pm 0.2$.  Note how the starless cores
almost all lie above the line and the  protostellar cores almost all
lie below the line, indicating that starless cores are in general more
flattened than protostellar cores.}
\label{fig:compare}
\end{figure}

To test whether the apparent difference between starless  cores and
protostellar cores is due to selection effects, we examined  whether
$q$ correlates with properties such as the linear size and  density of
a  core. In the Jijina et al. data set, the protostellar cores are
generally larger than the starless cores.  We believe this to be a
selection effect, since a protostellar core  is on average hotter and
more luminous than a starless core of the same  mass, and therefore
the outer parts of its envelope are more readily  detected. However
$q$ is uncorrelated with other properties (in  particular linear size)
and so we do  not believe that selection effects are biasing our
results significantly.

If selection effects cannot explain our result, then it is surprising. 
As explained by Zel'dovich (1970), and verified by many others since, 
one would expect a dynamically collapsing triaxial ellipsoid to collapse 
faster along its shortest axis than along the other axes. Consequently, 
more evolved cores (those with protostars) might be expected to have 
more extreme axial ratios than less evolved ones (starless cores), 
which is the opposite of our result. One possible explanation is that 
the collapse of cores with significant magnetic fields is not as 
simple as a pure gravitational collapse and may cause the effect we 
observe (eg.  Habe et al. 1991). Another possibility is that the outer 
envelopes of protostellar cores are not collapsing and have greater 
{\it isotropic} support than those of starless cores. For instance, 
they could have higher turbulence, as inferred from the broader 
line-widths observed by Beichman et al. (1986). We are currently 
undertaking a programme of simulations of the collapse of triaxial 
cores, which may shed light on the evolution of the axial ratios.

\begin{table}
\begin{center}
\begin{tabular}{ccccc} \hline
Core sample  & $\bar{\zeta}$ & $\sigma_{\zeta}$ & $\bar{\eta}$ &
$\sigma_{\eta}$ \\ \hline
All          & 0.8 & 0.1 & 0.4 & 0.2 \\ 
Starless     & 0.8 & 0.1 & 0.3 & 0.2 \\ 
Protostellar & 0.8 & 0.1 & 0.5 & 0.2 \\ \hline
\end{tabular}
\caption{The best fit triaxialities for mid- and short-axes for  the
starless and protostellar cores.}
\end{center}
\label{table:data}
\end{table}

\section{Conclusions}

We have shown that the projected axial ratios of molecular cores
(Jijina  et al. 1999) can be well fitted with randomly oriented
ellipsoids having their intrinsic axial ratios distributed according
to $\bar{\zeta} \simeq  0.8\,$, $\bar{\eta} \simeq 0.4\,$ and
$\sigma_\zeta \simeq \sigma_\eta  \simeq 0.1\,$. This conclusion is
different to that of Jones et al. (2001) in that they
infer less extreme axial ratios. The reason  for this is that they
allow distributions which (in effect) predict  no low $q$-values even
though several are observed.

When we divide the Jijina et al. (1999) data into subsamples and
repeat  our analysis we find a difference between starless  cores and
protostellar cores, in the sense that starless cores are flatter.
Specifically, starless cores are best fit with $\bar{\eta} \simeq
0.3\,$,  while protostellar cores are best fit with $\bar{\eta} \simeq
0.5\,$.  This result is surprising. A triaxial ellipsoid in free-fall
should collapse faster along its shortest axis than along the other
axes.  Hence older cores (those with protostars) might be expected to
have more extreme  axial ratios. It appears that the opposite is the
case.

%\section*{Acknowledgments}

\end{document}